\documentclass[letter]{aa}
\usepackage{graphicx}

\titlerunning{}

\def\Mo{M_{\odot}}
\def\Lo{L_{\odot}}

\begin{document}

\title{The Baade-Wesselink p-factor applicable to LMC Cepheids\thanks{Based on observations made with ESO telescopes
at the Silla Paranal Observatory under programme ID 280.A-5018(A) }}

\titlerunning{The projection factor of LMC Cepheids}
\authorrunning{Nardetto et al. }

\author{N. Nardetto \inst{1}, A. Fokin \inst{1,2}, P. Fouqu\'e\inst{3}, J. Storm\inst{4}, W. Gieren\inst{5}, G.
Pietrzynski \inst{5,6}, D. Mourard\inst{1}, P. Kervella\inst{7}}

\institute{Laboratoire Fizeau, UNS/OCA/CNRS UMR6525, Parc Valrose, 06108 Nice Cedex 2, France \\ 
\email{Nicolas.Nardetto@oca.eu}  \and Institute of Astronomy of the Russian Academy of Sciences, 48 Pjatnitskaya Str., Moscow 109017 Russia  \and Observatoire Midi-Pyr\'en\'ees, Laboratoire d'Astrophysique, UMR 5572, Universit\'e Paul Sabatier - Toulouse 3, 14 avenue Edouard Belin, 31400 Toulouse, France \and Leibniz-Institut f\"ur Astrophysik Potsdam (AIP), An der Sternwarte 16, D-14482 Potsdam, Germany \and Departamento de Astronom\'ia, Universidad de Concepci\'on, Casilla 160-C, Concepci\'on, Chile \and Warsaw University Observatory, Al. Ujazdowskie 4, 00-478, Warsaw, Poland \and LESIA, Observatoire de Paris, CNRS\,UMR\,8109, UPMC, Universit\'e Paris Diderot, 5 place Jules Janssen, 92195 Meudon, France}

\date{Received ... ; accepted ...}

\abstract{Recent observations of LMC Cepheids bring new constraints on the slope of the period-projection factor relation (hereafter \emph{Pp} relation) that is currently used in the Baade-Wesselink (hereafter BW)  method of distance determination. The discrepancy between observations and theoretical analysis is particularly significant for short period Cepheids} {We investigate three physical effects that might possibly explain this discrepancy: (1) the spectroscopic S/N that is systematically lower for LMC Cepheids (around 10) compared to Galactic ones (up to 300),  (2) the impact of the metallicity on the dynamical structure of LMC Cepheids, and (3) the combination of infrared photometry/interferometry with optical spectroscopy.} {To study the S/N we use a very simple toy model of Cepheids. The impact of metallicity on the projection factor is based on the hydrodynamical model of $\delta$ Cep already described in previous studies. This model is also used to derive the position of the optical versus infrared photospheric layers.} {We find no significant effect of S/N, metallicity, and optical-versus-infrared observations on the \emph{Pp}  relation.} {The \emph{Pp} relation of Cepheids in the LMC does not differ from the Galactic relation. This allows its universal application to determine distances to extragalactic Cepheids via BW analysis.}

\keywords{Stars: oscillations (including pulsations) }

\maketitle

\section{Introduction}\label{s_Introduction}
For decades the Cepheid stars have been used to calibrate the distance scale and the Hubble constant through their well-known period-luminosity (\emph{PL}) relation (Riess et al. 2009a, 2009b, 2011, and Freedman et al. 2010 for a review). Recently, using the Baade-Wesselink ($BW$) method to determine distances of Cepheids, Storm et al. (2011b, second paper of the series) has found that the $K$-band \emph{PL} relation is nearly universal and can be applied to any host galaxy regardless of its metallicity.

The projection factor is a key quantity of the BW methods: it is used to convert the radial velocity into the pulsation velocity of the star.  In their first paper Storm et al. (2011a) {\it directly} constrained the period-projection factor (\emph{Pp}) relation from observations. The zero point is based on the HST trigonometric parallaxes of Galactic Cepheids (Benedict et al. 2007, Groenewegen 2007, Fouqu\'e et al. 2007, M\'erand et al. 2004), and the slope is derived from BW distances of LMC Cepheids (all Cepheids in the LMC used by Storm et al. can be assumed to be at the same distance, leading to an extra constraint on the projection factor relation). Their relation is at 2~sigmas of the semi-theoretical relation of Nardetto et al.~(2007, 2009). For instance for $\delta$~Cep, they find $p=1.41\pm0.05$, while the theoretical value is $p=1.25\pm 0.05$ (Nardetto et al. 2009). An effect of the physical nature of LMC Cepheids on the slope of the \emph{Pp} relation is not excluded and should be investigated. This question of the universality of the \emph{Pp} relation is critical because it is a common assumption in the BW analysis of extragalactic Cepheids.

The spectroscopic S/N is systematically lower for LMC Cepheids (around 10) than Galactic ones (up to 300). In Sect.~1, we investigate the impact of S/N on the radial velocity, hence on the projection factor. In Sect.~2, we determine whether the projection factor should be corrected when observing LMC Cepheids of lower metallicity. Then, when combining an {\it optical} spectroscopic determination of the projection factor with an angular diameter derived from {\it infrared} observations, a mismatch of the position of the photosphere is not excluded. We investigate this possibility in Sect.~3. using new CRIRES spectroscopic observations and hydrodynamical modelling.


\section{S/N and p-factor}

\begin{figure}[htbp]
\begin{center}
\resizebox{0.84\hsize}{!}{\includegraphics[clip=true]{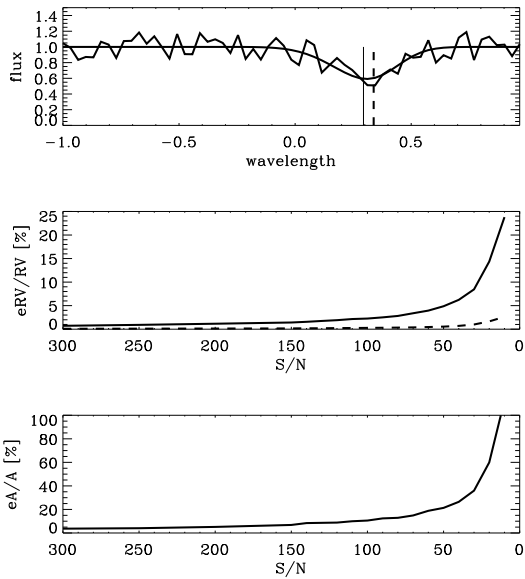}}
\end{center}
\caption{(a) The modelled random spectral line profile derived from our toy model (with $V_{\mathrm{rad}}=20$ km/s) together with the degraded profile (S/N of 10). The vertical lines correspond to the centroid (solid) and Gaussian (dashed) radial velocities. One hundred random spectral lines are calculated. (b) The relative uncertainties of the radial velocities as a function of the S/N (same legend). (c) The relative uncertainty on the bi-Gaussian line asymmetry as a function of the S/N.}  \label{Fig_geo} 
\end{figure}

The amplitude of the radial velocity curve depends on the method used to derive the position of the spectral line profile. Two main methods are currently used: the Gaussian fit and the first moment (or centroid). Only the first moment method is insensitive to the width of the spectral line (turbulence and/or stellar rotation) and is thus appropriate for comparing the dynamical structure of Cepheid's atmosphere (Burki et al. 1982, Nardetto et al. 2006). However, these two methods are not suitable for LMC Cepheids with spectra of very low S/N (around 10). The method currently used for distant (and thus faint) Cepheids is to cross-correlate thousands of spectral lines, derive a residual spectral line and apply the Gaussian fit method to determine the radial velocity (Baranne et al. 1986). Compared to a single line measurement of the radial velocity (using the first moment method), the projection factor is actually reduced (regardless of the period of the Cepheid) by about 5\% when using the cross-correlation method (Nardetto et al. 2009). The S/N might affect these different methods of radial velocity determination. 

We want to verify this assumption using a toy model already described in Nardetto et al. (2006). The modelled spectral line profile (Fig. 1a) is randomly degraded from an S/N of 300 to 10, by steps of 10. This is done one hundred times. For each S/N, we derive the mean radial velocity corresponding to the Gaussian and centroid methods (averaged over the 100 profiles), together with their respective mean uncertainties. A first result is that the mean radial velocity derived from both methods (Gaussian fit and first moment) is insensitive to the S/N. This is a good indication that the cross-correlation method is probably not biased by the S/N. 

Additional results have to be mentioned. The first moment method can lead to statistical uncertainties as large as 20\% when the S/N is lower around 10 (Fig.~1b). The Gaussian fit method is more robust with a few percent precision for an S/N of 10. One should, however, consider that these results  depend mainly on the radial velocity considered (i.e. the pulsation phase considered) and the spectral line depth. When using the cross-correlation method, the relative uncertainty on the radial velocity can drop to less than 1\%. We also take the opportunity to show that line asymmetry (derived using the bi-Gaussian method described in Nardetto et al. 2006) is highly sensitive to the S/N (Fig.~1c).  While the mean value of the spectral asymmetry is stable with the S/N (as found for the radial velocities), an S/N of 100 is required to derive a line asymmetry at a 10\% relative precision level. Studying the dynamical structure of pulsating stars atmospheres thus requires a minimum S/N of 100.  The conclusion of this section is that the S/N should not affect the period-projection factor relation of Galactic versus LMC Cepheids. 

\section{Metallicity and p-factor}

\begin{figure}[htbp]
\begin{center}
\resizebox{1.02\hsize}{!}{\includegraphics[clip=true]{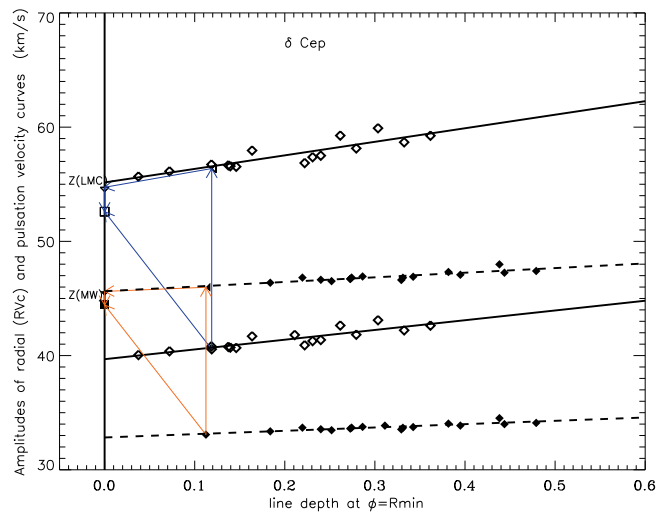}}
\end{center}
\caption{The decomposition of the projection factor. The filled and open diamonds correspond to the Milky Way and LMC models of $\delta$ Cep. For a given model, the lower (resp. higher) values are the amplitudes of the radial (resp. pulsation) velocity curves for each spectral line considered. The red and blue arrows indicate the decomposition of the projection factor for the MW and LMC respectively: $\uparrow$~($p_\mathrm{0}$), $\leftarrow$~($f_\mathrm{grad}$), $\downarrow$~($f_\mathrm{og}$) and $\nwarrow$~($p=p_\mathrm{0}f_\mathrm{grad}f_\mathrm{og}$). For detailed explanations, refer to Nardetto et al. (2007).} \label{Figuv}
\end{figure}

The metallicity of Cepheids has an impact on their $\kappa$-mechanism that drives the pulsation through an opacity process. As a consequence, the amplitude of the pulsation velocity of the star is affected by the metallicity: The lower the metallicity, the larger the amplitude of the pulsation velocity (Klagyivik \& Szabados 2007; Klagyivik \& Szabados 2009). But in principle, if the amplitude of the {\it pulsation} velocity is higher, the {\it radial} velocity should also be larger in the same proportion, and the projection factor $p=\frac{V_{\mathrm{puls}}}{V_{\mathrm{rad}}}$ should {\it not} change. The key-point is then to know whether the velocity gradient in the atmosphere of the star changes with the metallicity in such a way that the projection factor for a given spectral line will change. To answer this question, we consider the hydrodynamical model of $\delta$~Cep in the Milky Way (Nardetto et al.~2004), based on the code by Fokin et al. (2004), and compare it to a consistent LMC model of the same star. The fundamental parameters of the LMC model of $\delta$ Cep are determined as follows. Following  Luck et al.~(1998) and Romaniello et al.~(2008), we first assume $Z=0.008$ (very close to their respective values of 0.01 and 0.009). We then consider a helium-to-metal enrichment ratio $\frac{\Delta Y}{\Delta Z}$ of $1.5$ (consistent with the value of $2.1\pm0.9$ derived by Casagrande et al.~2007), which leads to $X=0.730$ and $Y=0.262$. Then, we assume the same mass for $\delta$~Cep in the Milky Way and the LMC (i.e. $M=4.8\Mo$) and we simply use the mass-luminosity relation of Chiosi et al.~(1993, Eq. 25) to derive the luminosity of $\delta$~Cep in the LMC. We obtain $L=2290\Lo$. Finally, using the Milky Way and LMC analytical relations of the red edge instability strip by Bono et al.~(2000, Tab. 5), we find an expected effective temperature of about $T_{\mathrm{eff}}= 6100K$. All parameters used in the code are presented in Table 1 for the Galactic and LMC models of $\delta$~Cep. We find a pulsation period of $P=5.38$d (in LMC), which is consistent with theoretical results by Bono et al. (2000, Table 6). 

For both models (with Galactic and LMC metallicities), we calculated the spectral line profiles for the 17 metallic lines presented in Nardetto et al. (2007, hereafter Paper I) using radiative transfer equations in a moving atmosphere. The calculation is made at 40 equidistant pulsation phases. We finally plotted the amplitude of the {\it radial} (using the first moment method) and {\it pulsation} velocity curves as a function of the line depth at minimum radius (see Fig.~2) in order to measure the velocity gradient and perform the decomposition of the projection factor as explained in Paper~I. There are three subconcepts involved in the decomposition of the projection factor: 1/~the geometric projection factor mainly related to the limb-darkening of the star ($p_\mathrm{0}$), 2/~the correction due to the velocity gradient within the atmosphere ($f_\mathrm{grad}$) and 3/~the correction due to the relative motion between the {\it optical} and {\it gas} layers corresponding to the photosphere ($f_\mathrm{og}$); for a detailed definition of these two layers, refer to Eqs. 1 and 3 of Nardetto et al. (2004). This decomposition of the projection factor for the Milky Way and the LMC models of $\delta$~Cep are indicated in Fig.~2, and the corresponding values can be found in Table 1. To compare the MW and LMC projection factors properly, we have to consider two spectral lines of similar depth (around 0.1):  FeI 4896.439\AA\ for the MW and FeI 5373.709\AA\ for the LMC. These choices only have an impact on $f_\mathrm{grad}$.  

We find that the projection factor is about  the same for the Galactic and LMC models (at a 1.5\% level). This basically means that even if the amplitude of the radial and pulsation velocity curves are larger for the LMC metallicity model, the structure of the atmosphere is relatively unchanged:  the velocity gradient is larger (in absolute terms), but it is the same relative to the amplitude of the velocities. As a conclusion, the p-factor value appropriate for a Cepheid of a given period should not depend on its metallicity, and the large discrepancy obtained for the \emph{Pp}ñ relations (observation versus theory), especially for short-period Cepheids (for  $\delta$ Cep, $p=1.41$[observations] versus $p=1.25$[theory]), is not explained by a difference of metallicity. 

\begin{table}
\begin{center}
\caption[]{Hydrodynamical models and corresponding p-factors}
\label{Tab_hydro}
\begin{tabular}{ccccc}
\hline \hline \noalign{\smallskip}
parameters                                      &     MW    &     LMC     \\ 
\hline
$L$  [$\Lo$] &  $1995$ & $2290$  \\ 
$M$  [$\Mo$] &  $4.8$ & $4.8$  \\ 
$T_{\mathrm{eff}}$  [K] &  $5877$ & $6100$  \\ 
$X$                 &$ 0.700$ & $0.730 $  \\
$Y$                & $0.280$ & $0.262$  \\
$Z$                & $0.020$ & $0.008$  \\
$P$  [days]              & $5.41$ & $5.38$  \\
\hline
$p_{\mathrm{0}}$ & $1.390 $ & $1.390$  \\
$f_{\mathrm{grad}}$ & $0.990 $ & $0.975$  \\
line for $f_{\mathrm{grad}}$ [\AA]  & FeI 4896 & FeI 5373  \\
$f_{\mathrm{og}}$ & $0.963 $ & $0.963$  \\
$p=p_{\mathrm{0}}f_{\mathrm{grad}}f_{\mathrm{og}}$ & $1.325 $ & $1.305$   \\

\hline \noalign{\smallskip}
\end{tabular}
\end{center}
\end{table}

\section{Optical p-factor versus infrared diameters}

When applying the BW method, infrared photometry/interferometry is often combined with optical spectroscopy to derive the distance of the star. In principle, the radial velocity curve derived from {\it optical} spectroscopy has to be combined with an {\it optical} projection factor to obtain the pulsation velocity associated to the {\it optical} photospheric layer. As a consequence, to get an accurate determination of the distance, the limb-darkened angular diameter derived from the infrared surface brightness method or from infrared long-baseline interferometry must correspond to the same {\it optical} photospheric layer. In other words, an implicit assumption in the BW method is to consider that the photospheric layer in the {\it optical} and in the {\it infrared} are associated with the same material in the star. However, the larger the wavelength, the lower the position of the photosphere in the star (see theoretical results by Sasselov et al. 1990, Fig.~21~and~22). This effect seems to be stronger for long-period Cepheids (their Fig. 21), than for short-period Cepheids (their Fig. 22). Verifying these theoretical results with observations is extremely difficult. The best is probably to use interferometry. From the spectrocopic point of view, one can compare the optical and infrared radial velocity curves of two spectral lines of similar depths (using the first moment method). However, the velocity ratio obtained has to be corrected from the velocity gradient and from optical-versus-gas layer effect ($f_\mathrm{grad}$ and $f_\mathrm{og}$, respectively). For instance, it is probably not possible to perform this analysis for atypical Cepheids, like X Sgr presenting a double shock wave in the atmosphere (Mathias et al. 2006), as done by Sasselov et al. (1990).

We present infrared CRIRES observations of $\ell$~Car obtained in 2008-2009 at four different pulsation phases (S/N of about 300 in the continuum).  We find three unblended spectral lines, Br$\gamma$ 21661.2\AA\  and two metallic lines:  FeII 21266.1\AA\  and CaI 22090.0\AA. In Fig.~3, we present the CaI spectral line flux as a function of time. The first moment radial velocity and depth (as a percentage of continuum) are indicated in Fig.~4, together with HARPS observations of a spectral line of similar depth.  The dashed line in Fig.~4 (lower panel) is the optical HARPS radial velocity curve multiplied by 1.30, showing that the infrared radial velocity curve has an amplitude that is about 30\% larger than in the optical. The conclusion is that, even if these two metallic lines in the optical and in the infrared have a similar depth (which means in principle that their line-forming regions have a similar optical depths, and that they form approximatively at the same position in the atmosphere compared to the photosphere), the infrared amplitude of the radial velocity curve is significantly larger than the optical one. This suggests (as explained above) that the photospheric layer is probably forming higher in the infrared than in the optical (which seems to be in contradiction with Sasselov et al. 1990). More data are required to confirm this result.

We address this question theoretically using our hydrodynamical models of $\delta$~Cep and $\ell$~Car. The model of $\ell$ Car is described in Paper I. We consider the solar metallicity. Concerning $\delta$~Cep, the optical and infrared photospheric layers ($\tau_{\mathrm{c}}=1$) form at the same layer in the atmosphere, whereas for $\ell$ Car the infrared photosphere forms 2\% higher than the optical one. This implies a 2\% increase only in the projection factor, which seems to be in contradiction with Sasselov et al. (1990) and also CRIRES data. This delicate question will be studied in more detail later, using a larger sample of infrared (resp. optical) spectroscopic (resp. interferometric) data.  In either case, this effect (optical versus infrared) should affect the Milky Way and LMC Cepheids in the same way.   


\begin{figure}[htbp]
\begin{center}
\resizebox{0.85\hsize}{!}{\includegraphics[clip=true]{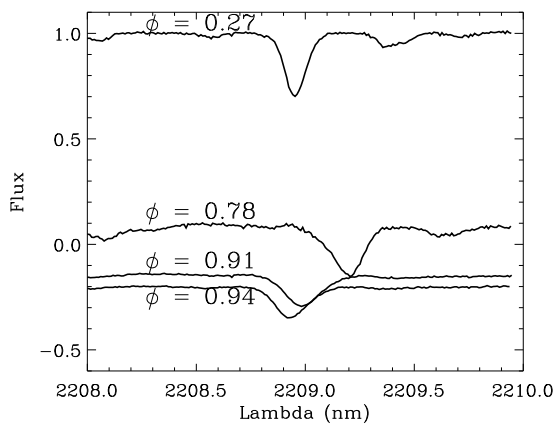}}
\end{center}
\caption{The CRIRES spectral line profiles (CaI 22090.0\AA) of $\ell$~Car as a function of the pulsation phase.} \label{Fig_crires}
\end{figure}
\begin{figure}[htbp]
\begin{center}
\resizebox{0.80\hsize}{!}{\includegraphics[clip=true]{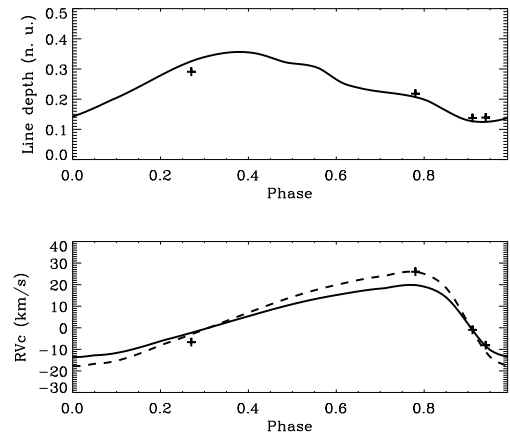}}
\end{center}
\caption{The line depth (up) and the first moment radial velocity (down) corresponding to the infrared CRIRES spectral line of $\ell$~Car (crosses) is compared to the optical 4896A HARPS spectral line of similar depth (solid line). The meaning of the dashed line is explained in the text.} \label{Fig_comparison}
\end{figure}
\vspace{-0.9cm}

\section{Conclusions}
We find that the S/N, metallicity, and optical-versus-infrared observations (but this last point has to be confirmed) cannot explain the discrepancy between the theoretical and empirical period-projection factor relation based on LMC observations. Other possibilities can be considered, such as limb darkening for instance. Indeed, following the relation by Storm et al. (2011a), the very short-period Cepheids should have a projection factor close to 1.45-1.5. This means a limb-darkening close to zero (uniform disk). By constraining this limb darkening using interferometry (for instance VEGA/CHARA in optical, Mourard et al. 2009), one can derive a geometric projection factor that should help in resolving the discrepancy. However, our results seem to indicate that the \emph{Pp} relation is universal. This is extremely precious when applying the BW method to extragalactic Cepheids.

\begin{acknowledgements}
WG and GP gratefully acknowledge financial support for this work from
the Chilean Center for Astrophysics FONDAP 15010003, and from the BASAL
Centro de Astrofisica y Tecnologias Afines (CATA) PFB-06/2007. Support from the Polish grant N203 387337 and TEAM subsidy of the Fundation for Polish Science (FNP) is also acknowledged. This research received the support of PHASE, the high angular resolution partnership between ONERA, Observatoire de Paris, CNRS and University Denis Diderot Paris 7. NN acknowledges F. Millour, J. Groh, and S. Kraus, who helped prepare the CRIRES DDT proposal, and L. Szabados for a careful reading of the paper. We finally thank the referee for his extremely useful comments. 
\end{acknowledgements}

\end{document}